\documentclass[preprintnumbers,superscriptaddress,pra,showkeys]{revtex4}
\usepackage{amsfonts}
\usepackage{amssymb}
\usepackage{amsmath}
\usepackage{epsfig}
\usepackage{graphicx}
\usepackage{color}

\input{tcilatex}

\begin{document}

\title{Geometrical aspect of susceptibility critical exponent}
\author{Q. H. Liu}
\email{quanhuiliu@gmail.com}
\affiliation{School for Theoretical Physics, School of Physics and Electronics, Hunan
University, Changsha 410082, China}
\date{\today }

\begin{abstract}
Critical exponent $\gamma \succeq 1.1$ characterizes behavior of the
mechanical susceptibility of a real fluid when temperature approaches the
critical one. It results in zero Gaussian curvature of the local shape of
the critical point on the thermodynamic equation of state surface, which
imposes a new constraint upon the construction of the potential equation of
state of the real fluid from the empirical data. All known empirical
equations of state suffer from a weakness that the Gaussian curvature of the
critical point is negative definite instead of zero.
\end{abstract}

\author{}
\maketitle

Geometry plays important roles in thermodynamics. The modern version of
thermodynamics can be reformulated in terms of contact geometry \cite{geom1}%
, and the so-called geometry of thermodynamics has been put forward which
describes the space of thermodynamic parameters by the Riemannian metric 
\cite{geom2,geom3,geom4}. The influence of the curved space on the critical
behavior of the two-dimensional Ising model is identified \cite{geom5}, and
geometric critical exponents are definable in classical and quantum phase
transitions \cite{geom6}. The new relationship between thermodynamical and
geometry is always interesting, and we report a new requirement on the
construction of the empirical equation of state (EoS) based on the
differential geometry of the surfaces.

The elaboration of a better form of the empirical EoS best fitting the
experimental data and also meeting the theoretical requirements have been an
important issue for more than one century \cite%
{wei,eos2,lw,eos1,review1,review2,exp}. For a real fluid, a theoretical
problem which remains open for a long time is, in the close neighborhood of
the critical point, what is the precise form of the EoS?\ For instance, it
is well-known that both the van der Waals EoS and the general theory of the
Landau theory of phase transitions, predicts the  susceptibility critical
exponent $\gamma =\gamma ^{\prime }$ to be $1$, and all known empirical EoS
fail to exactly reproduce the experimental values $\gamma =1.2\sim
1.3\gtrsim \gamma ^{\prime }=1.1$ $\sim 1.2$ \cite{toda,pathria}. Recently,
we conjecture that the Gaussian curvature of the local shape of the
vapor-liquid critical point is zero \cite{liu2021}. In present paper, we
prove that the conjecture is true, and secondly construct an fluid EoS which
has $\gamma =\gamma ^{\prime }=3$, which is compatible the zero Gaussian
curvature of the vapor-liquid critical point.

The fluid of a pure substance belongs to the so-called $pVT$ system, which
means that the EoS usually takes following form,%
\begin{equation}
p=p(V,T),  \label{eos}
\end{equation}%
where $p,V,T$ denote the pressure, volume and temperature, respectively. In
general, this function has continuous first and second order derivatives at
the critical point of which three parameters $\left(
p_{c},V_{c},T_{c}\right) $ satisfy, in addition to the EoS (\ref{eos}),%
\begin{equation}
\left( \frac{\partial p}{\partial V}\right) _{T_{c}}=0,\left( \frac{\partial
^{2}p}{\partial V^{2}}\right) _{T_{c}}=0.  \label{cp}
\end{equation}%
In present paper, we do not deal with piecewise or other discontinuous form
of EoS (\ref{eos}).

To note that in geometry the EoS (\ref{eos}) can be viewed as a
two-dimensional surface in the flat $pVT$ space, and its shape can be
completely characterized by the mean and Gaussian curvature. \cite{docarmo}
It is then interesting to explore the local shape of the vapor-liquid
critical point via these two curvatures. In calculation, the dimensionless
EoS surface equation (\ref{eos}) must be used, in which all quantities $%
\left( p,V,T\right) $ are transformed into those referring to units $\left(
p_{c},V_{c},T_{c}\right) $, or other units $\left( p^{\prime },V^{\prime
},T^{\prime }\right) $ of specific states, respectively. The transformed
form of EoS bears a resemblance to the law of corresponding states of the
van der Waals EoS. The mean curvature $H$ and Gaussian curvature $K$ are,
respectively, \cite{docarmo} 
\begin{eqnarray}
H &=&\frac{\left( \frac{\partial ^{2}p}{\partial V^{2}}\right) _{T}\left(
\left( \frac{\partial p}{\partial T}\right) _{V}^{2}+1\right) +\left( \frac{%
\partial ^{2}p}{\partial T^{2}}\right) _{V}\left( \left( \frac{\partial p}{%
\partial V}\right) _{T}^{2}+1\right) -2\left( \frac{\partial p}{\partial V}%
\right) _{T}\left( \frac{\partial p}{\partial T}\right) _{V}\left( \frac{%
\partial ^{2}p}{\partial V\partial T}\right) }{2\left( \left( \frac{\partial
p}{\partial V}\right) _{T}^{2}+\left( \frac{\partial p}{\partial T}\right)
_{V}^{2}+1\right) ^{3/2}},  \label{GenH} \\
K &=&\frac{\left( \frac{\partial ^{2}p}{\partial V^{2}}\right) _{T}\left( 
\frac{\partial ^{2}p}{\partial T^{2}}\right) _{V}-\left( \frac{\partial ^{2}p%
}{\partial V\partial T}\right) ^{2}}{\left( \left( \frac{\partial p}{%
\partial V}\right) _{T}^{2}+\left( \frac{\partial p}{\partial T}\right)
_{V}^{2}+1\right) ^{2}}.  \label{GenK}
\end{eqnarray}%
At the critical point the conditions (\ref{cp})$\ $apply, we have the mean
curvature $H_{C}$ and Gaussian curvature $K_{C}$, respectively, 
\begin{equation}
H_{C}=\frac{\left( \frac{\partial ^{2}p}{\partial T^{2}}\right) _{V}}{%
2\left( \left( \frac{\partial p}{\partial T}\right) _{V}^{2}+1\right) ^{3/2}}%
,K_{C}=-\frac{\left( \frac{\partial ^{2}p}{\partial V\partial T}\right) ^{2}%
}{\left( \left( \frac{\partial p}{\partial T}\right) _{V}^{2}+1\right) ^{2}}.
\label{CHK}
\end{equation}

The compressibility or a mechanical response function 
\begin{equation}
\kappa _{T}\equiv -\frac{1}{V}\left( \frac{\partial V}{\partial p}\right)
_{T}
\end{equation}%
corresponds to a susceptibility%
\begin{equation}
\left( \frac{\partial p}{\partial V}\right) _{T}=-\frac{1}{V\kappa _{T}}.
\end{equation}%
Near the critical point, the experiments suggest, with $\gamma =\gamma
^{\prime }\succeq 1.1$, \cite{toda,pathria} 
\begin{equation}
\kappa _{T}\propto \left\{ 
\begin{array}{cc}
\left( T-T_{c}\right) ^{-\gamma }, & \left( T\rightarrow T_{c}+0\right)  \\ 
\left( T_{c}-T\right) ^{-\gamma ^{\prime }}, & \left( T\rightarrow
T_{c}-0\right) 
\end{array}%
\right. .
\end{equation}%
In consequence, we have, 
\begin{equation}
\frac{\partial ^{2}p}{\partial V\partial T}=-\frac{1}{V}\frac{\partial }{%
\partial T}\left( \frac{1}{\kappa _{T}}\right) \propto \left\{ 
\begin{array}{cc}
\left( T-T_{c}\right) ^{\gamma -1}, & \left( T\rightarrow T_{c}+0\right)  \\ 
\left( T_{c}-T\right) ^{\gamma ^{\prime }-1}, & \left( T\rightarrow
T_{c}-0\right) 
\end{array}%
\right. =0,
\end{equation}%
which can be rewritten into, in terms of the Gaussian curvature from (\ref%
{CHK}), 
\begin{equation}
K_{C}=0.  \label{kc}
\end{equation}%
Thus, for real fluids, the Gaussian curvature of local shape of the critical
point of the EoS surface is zero.

In contrast, none of the known typical empirical EoS can reproduce $K_{C}=0$%
, whose results are listed in following Table I. In the last line of the
Table I, the Shamsundar-Lienhard EoS \cite{review2} is special, for it is
rather a principle than an explicit form of an equation. "The shape of the
(experimental data) figure tells us that a cubic-like equation must be of
the form" \cite{review2}, and "the advantage of this form is that it
automatically satisfies critical point criteria, ... ." \cite{review2}
Though these empirical EoS in the Table I do not exhaust all possibilities,
we are safe to say that $K_{C}=0$ is beyond the current form of the Landau
theory of phase transitions for it predicts $\gamma =\gamma ^{\prime }=1,$
thus $K_{C}\prec 0$. \cite{toda,pathria}

\begin{table}[th]
\caption{ Mean and Gaussian curvature of the critical points for empirical
or semi-empirical EoS. Symbol $v$ denotes the molar volume, $a$ and $b$ are
repulsive and attractive parameter respectively, and $y=b/4v$, and $a(T)$
depends on temperature as well as on other substance parameters \protect\cite%
{wei,exp}. Only one typical value of the substance parameter in $a(T)$ is
used in the calculations.}\centering
\par
\begin{tabular}{|l|l|l|l|l|l|}
\hline
No & References & EoS & $H_{C}$ & $K_{C}$ & Year \\ \hline
1 & van der Waals & $\frac{RT}{v-b}-\frac{a}{v^{2}}$ & 0 & -0.125 & 1873 \\ 
\hline
2 & Dieterici & $\frac{RT}{v-b}\exp \left( -\frac{a}{vkT}\right) $ & 0.063 & 
-0.04 & 1916 \\ \hline
3 & Redlich-Kwong & $\frac{RT}{v-b}-\frac{a}{T^{1/2}v(v+b)}$ & -0.006 & 
-0.065 & 1949 \\ \hline
4 & Thiele & $\frac{RT}{v}\frac{1+y+y^{2}}{\left( 1-y\right) ^{3}}-\frac{a}{%
v^{2}}$ & 0 & -0.099 & 1963 \\ \hline
5 & Guggenheim & $\frac{RT}{v}\frac{1}{1-y^{4}}-\frac{a}{v^{2}}$ & 0 & -0.101
& 1965 \\ \hline
6 & Carnahan-Starling & $\frac{RT}{v}\frac{1+y+y^{2}-y^{3}}{\left(
1-y\right) ^{3}}-\frac{a}{v^{2}}$ & 0 & -0.099 & 1969 \\ \hline
7 & Soave & $\frac{RT}{v-b}-\frac{a\left( T\right) }{v(v+b)}$ & -0.004 & 
-0.045 & 1972 \\ \hline
8 & Peng-Robinson & $\frac{RT}{v-b}-\frac{a\left( T\right) }{v(v+b)+b(v-b)}$
& -0.003 & -0.058 & 1976 \\ \hline
9 & Shamsundar-Lienhard & $g(T)\left( 1-\frac{\left( V-V_{l}\right) \left(
V-V_{m}\right) \left( V-V_{g}\right) }{f(V,T)}\right)$ \QQfnmark{%
In this EoS, symbols $g(T)$, $V_{l}$, $V_{m}$, $V_{g}$ stands for four
parameters depending on the temperatures, and $f(V,T)$ depends on both
volume and temperature and possesses no poles or roots in the physical range
of $V$. The explicit form of $f(V,T)$ depends on other three parameters,
details of which do not affect our conclusion.} & / & 0 & 1993 \\ \hline
\end{tabular}%
\QQfntext{0}{
In this EoS, symbols $g(T)$, $V_{l}$, $V_{m}$, $V_{g}$ stands for four
parameters depending on the temperatures, and $f(V,T)$ depends on both
volume and temperature and possesses no poles or roots in the physical range
of $V$. The explicit form of $f(V,T)$ depends on other three parameters,
details of which do not affect our conclusion.}
\end{table}

Now we present an EoS with $K_{C}=0$. Our constructed EoS takes following
form,%
\begin{equation}
p=\frac{8T}{9\left( V-\frac{1}{3}\left( \frac{19}{3}-\frac{6}{V}+\frac{2}{%
V^{2}}\right) \right) }-\frac{3\left( 3-\frac{3}{T}+\frac{1}{T^{2}}\right) }{%
V^{2}}  \label{liu}
\end{equation}%
where $\left( p,V,T\right) $ are written in unit $\left(
p_{c},V_{c},T_{c}\right) \left( =\left( 1,1,1\right) \right) $. It can be
considered a highly distorted version of law of corresponding state of the
van der Waals equation. The isotherms given by this EoS (\ref{liu}) are
plotted in Fig. 1. Below the critical temperature $T_{C}$, the isotherms
explicitly exhibit van der Waals loops. Straightforward calculations show
that both the mean and the Gaussian curvature at the critical point are
zero, i.e., $H_{C}=K_{C}=0$, and also an symmetrical susceptibility critical
exponent, 
\begin{equation}
\gamma =3\text{ }(T\rightarrow T_{C}+0)\text{, and }\gamma ^{\prime }=3\text{
}(T\rightarrow T_{C}-0).
\end{equation}%
It is quantitatively different from the susceptibility critical exponent for
the real fluid. The origin of the difference may lie in that we use the
continuous form of the EoS from which we have in fact attempted many times,
which shall be explored in the future. 

So, far, there are at least three requirements on the form of EoS for the
real fluid. 1) In large volume limit, the EoS reproduces the ideal gas law. 
\cite{review2} 2) At both ends of a vapor-liquid coexistence line, there are
onset and outset value for volume for gas and liquid state, and also a
coexisting pressure. \cite{review2} 3) The Gaussian curvature of the
critical point is zero. All these constraints directly come from
experiments. There may be other constraints on the form of the EoS, which
may arise from the theoretical requirements such as Maxwell area
construction for van der Waals loop, \cite{review2} which must used case by
case.

In summary, geometry not only offers better understanding and deeper insight
into the mathematical structure of the thermodynamics, but also presents an
accurate and convenient means to characterize various properties of
thermodynamic states. We report that the local shape of the vapor-liquid
critical point on thermodynamic surface has zero Gaussian curvature, which
has long hidden in the susceptibility critical exponent $\gamma \succ 1$. It
can be used to distinguish different empirical models, and to impose on the
construction of the new EoS as well. A new form of the EoS capturing the
essential feature of the vapor-liquid phase transition with $\gamma \succ 1$
is successfully elaborated, which is still quantitatively different from the
susceptibility critical exponent for the real fluid. 

\begin{acknowledgments}
This work is financially supported by National Natural Science Foundation of
China under Grant No. 11675051.
\end{acknowledgments}

\bigskip

\begin{figure}[h]
\includegraphics[height=9cm]{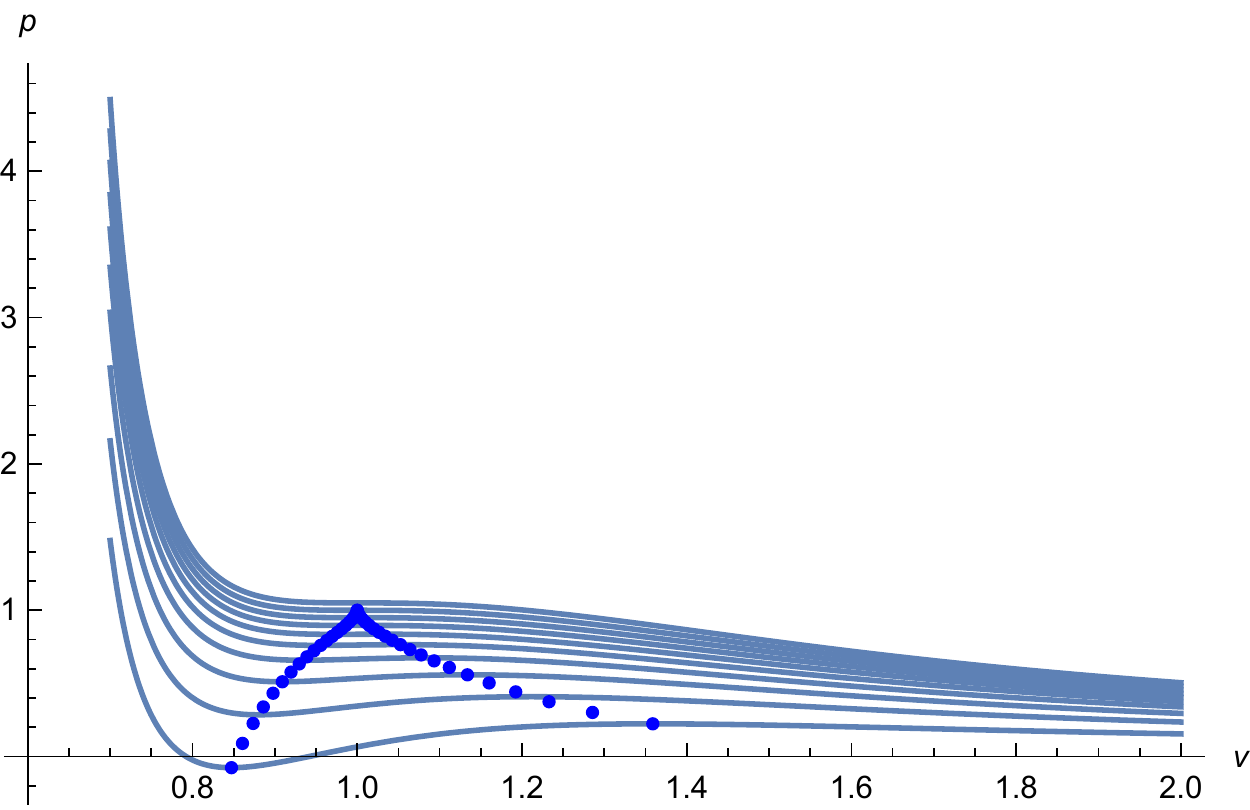}
\caption{The isotherms (solid lines) with $T=0.60,0.65,0.70,...,1.05$ from
undermost to upmost, and the spinodal curve (solid dots) for EoS (\protect
\ref{liu}). It clearly shows that the EoS captures the essential feature of
the vapor-liquid phase transition.}
\label{Fig1}
\end{figure}


\begin{thebibliography}{99}
\bibitem{geom1} R. Hermann, Geometry, Physics and Systems (New York: Dekker
1973).

\bibitem{geom2} F. Weinhold, Thermodynamics and geometry. \textit{Phys. Today%
} \textbf{29}(1976)23-30.

\bibitem{geom3} R. Gilmore, Length and curvature in the geometry of
thermodynamics, \textit{Phys. Rev. A} \textbf{30}(1984)1994.

\bibitem{geom4} G. Ruppeiner, Riemannian geometry in thermodynamic
fluctuation theory, \textit{Rev. Mod. Phys.} \textbf{67}(1995)605-659.

\bibitem{geom5} H. Shima and Y. Sakaniwa, Geometric effects on critical
behaviours of the Ising model, \textit{J. Phys. A: Math. Gen.} \textbf{39}%
(2006)4921.

\bibitem{geom6} P. Kumar and T. Sarkar, Geometric critical exponents in
classical and quantum phase transitions, \textit{Phys. Rev. E} \textbf{90}%
(2014)042145.

\bibitem{wei} Y. S. Wei and R. J. Sadus, Equations of State for the
Calculation of Fluid-Phase Equilibria, \textit{AIChE Journal,} \textbf{46}%
(2000)169

\bibitem{eos2} G. M. Kontogeorgis, R. Privat, and Jean-Noel Jaubert, Taking
Another Look at the van der Waals EoS -- Almost 150 Years Later. \textit{J.
Chem. Eng. Data}, \textbf{64}(2019)4619-4637.

\bibitem{lw} J. L. Lebowitz, E. M. Waisman, Statistical Mechanics of Simple
Fluids: Beyond van Der Waals. \textit{Phys. Today}, \textbf{33}(1980)24-30.

\bibitem{eos1} R. C. Reid, J. M. Prausnitz and B. E. Poling, The Properties
of Gases and Liquids (New York: McGraw-Hill, 1987).

\bibitem{review1} J. H. Lienhard, N. Shamsundar and P.O. Biney, Spinodal
Lines and Equations of State-a review, \textit{Nucl. Eng. Des.} \textbf{95}%
(1986)297-313.

\bibitem{review2} N. Shamsundar and John H. Lienhard, Equations of state and
spinodal lines a review, \textit{Nucl. Eng. Des.} \textbf{141}(1993)269-287.

\bibitem{exp} A. Anderko, Cubic and generalized van der Waals equations, 
\textit{Experimental Thermodynamics}, \textbf{5}( 2000)75-126.

\bibitem{toda} M. Toda, R. Kubo, N. Saito, \textit{Statistical Physics I:
Equilibrium Statistical Mechanics}, 2nd Ed., (Berlin: Sringer-Verlag, 2012).

\bibitem{pathria} R. K. Pathria, P. D. Beale, \textit{Statistical Mechanics}%
, 3rd ed., (Oxford: Butterworth-Heinemann, 2011).

\bibitem{liu2021} J. S. Yu, X. Zhou, J. F. Chen, W. K. Du, X. Wang and Q. H.
Liu, Local Shape of the Vapor--Liquid Critical Point on the Thermodynamic
Surface and the van der Waals EoS. \textit{Front. Phys.} \textbf{9}%
(2021)679083.

\bibitem{docarmo} M. P. do Carmo, \textit{Differential Geometry of Curves
and Surfaces} (New York: Prentice-Hall, 1976).

\bibitem{bose} J. D. Gunton and M. J. Buckingham, Condensation of the Ideal
Bose Gas as a Cooperative Transition, \textit{Phys. Rev.} \textbf{166}%
(1968)152.
\end{thebibliography}
\end{document}